\documentclass[twocolumn,aps,showpacs,prb]{revtex4-1}
\usepackage{graphicx}
\usepackage{epsfig}

\begin{document}

\title{Model of ultrafast demagnetization driven by spin-orbit coupling
in a photoexcited antiferromagnetic insulator Cr$_{2}$O$_{3}$}

\author{ Feng Guo, Na Zhang, Wei Jin, Jun Chang}
\email{jun.chang@hotmail.com}

\affiliation{College of Physics and Information Technology, Shaanxi Normal
University, Xi'an 710119, China~
}


\begin{abstract}
We theoretically study the dynamic time evolution following laser pulse
pumping in an antiferromagnetic insulator Cr$_{2}$O$_{3}$. From the photoexcited high-spin quartet states to the long-lived low-spin doublet states, the ultrafast demagnetization processes are investigated by solving the dissipative Schr\"odinger equation. We find that  the demagnetization times are of the order of hundreds of femtosecond, in good agreement with recent experiments. The switching times could be strongly reduced by properly tuning the energy gaps between the multiplet energy levels of Cr$^{3+}$. Furthermore, the relaxation times also depend on the hybridization of atomic orbitals in the first photoexcited state. Our results suggest that the selective manipulation
of electronic structure by engineering stress-strain or chemical substitution
allows effective control of the magnetic state switching in photoexcited
insulating transition-metal oxides. 
\end{abstract}

\maketitle

\section*{Introduction}

In recent years, growing attention has been drawn to the photodriven
ultrafast control of the quantum states and the physical properties
in solid-state and molecular systems. In addition to the great theoretical interest in understanding
the nonequilibrium dynamics in materials, it could be applied technically to
the magnetic or electronic recording. \cite{Sato2016Dynamic,Cammarata2014,Jin2014,Kirilyuk2010Ultrafast,Collet2003Laser}
The photoinduced change of physical properties is often attributed to thermal effects because the photon energy eventually is redistributed
among interacting charge, spin, and lattice degrees of freedom, and
increases the system temperature instantly. \cite{Beaurepaire1996}
On the other hand, photoirradiation may induce non-thermal metastable
states or transient phases with optical, magnetic and electric properties
distinct from that of the ground states. \cite{Zhang2015,Ehrke2011,Ichikawa2011}

Among these light-responsive materials, the ferromagnetic materials
have been brought into sharp focus by laser-induced demagnetization
since Bigot and coworkers found the ultrafast dropping of magnetization
in nickel film following optical pulses in 1996. \cite{Beaurepaire1996} Until recently, the ultrashort pulses of light are applied to manipulate
the ultrafast processes in the antiferromagnets. \cite{Marti2014Room,Loth2012Bistability,Forst2011Driving,Fiebig2008Ultrafast,Kimel2004Laser}
Indeed, antiferromagnetic (AFM) materials have more advantages than ferromagnets. For example, they are insensitive to external magnetic fields, \cite{Marti2014Room}
stable in miniaturization \cite{Loth2012Bistability} and much faster
in controlling spin dynamics. \cite{Kimel2004Laser}

AFM insulator Chromium oxide (Cr$_{2}$O$_{3}$) has been the subject
of study since the 1960s and its electronic and static optical properties
are now well understood. \cite{Mcclure1963Comparison,Macfarlane1963Analysis,Mo1997Modulation,Allos1977Ultraviolet,Forster2004Excited,Dodge1999Time,Muthukumar1996Theory,Tanabe1998Interference,Wang1998The}
However, the ultrafast dynamic demagnetization processes were not
probed until recently. \cite{Sala2016Resonant,Satoh2007Ultrafast,Lefkidis2009,Fiebig2008Ultrafast}
The time-resolved second harmonic generation is applied to probe the
time evolution of the magnetic and structural state following laser illuminations in the AFM insulator. Variations in the pump photon-energy
lead to either localized transitions within the metal-centered states of the Cr ion or charge transfer
between Cr and O. Despite its relevance to industrial technology,
the ultrafast processes of demagnetization are not well understood
from quantum nonequilibrilium dynamics. To selectively control the
demagnetization rate is still at a tentative stage in experiments.

In this paper, we first construct a local quantum-mechanical demagnetization
model of the photoinduced electron states in Cr$_{2}$O$_{3}$. The
effects of the energy dissipations are taken into account by a dissipative
Schr\"odinger equation. We simulate the time evolution of the excited
states following 1.8 eV and 3.0 eV light illumination and find that the
decay times from the high-spin quartet states to the low-spin doublet
states range from 300 femtoseconds (fs) to 450 fs, in line with the experiments. \cite{Sala2016Resonant} We show that the ratio of the energy gap to
the electron-phonon self-energy has a marked impact on the demagnetization
times. The decay times are also influenced by the hybridization of
atomic orbitals in the first photoexcited state.

\begin{figure}[t]
\includegraphics[width=1\columnwidth]{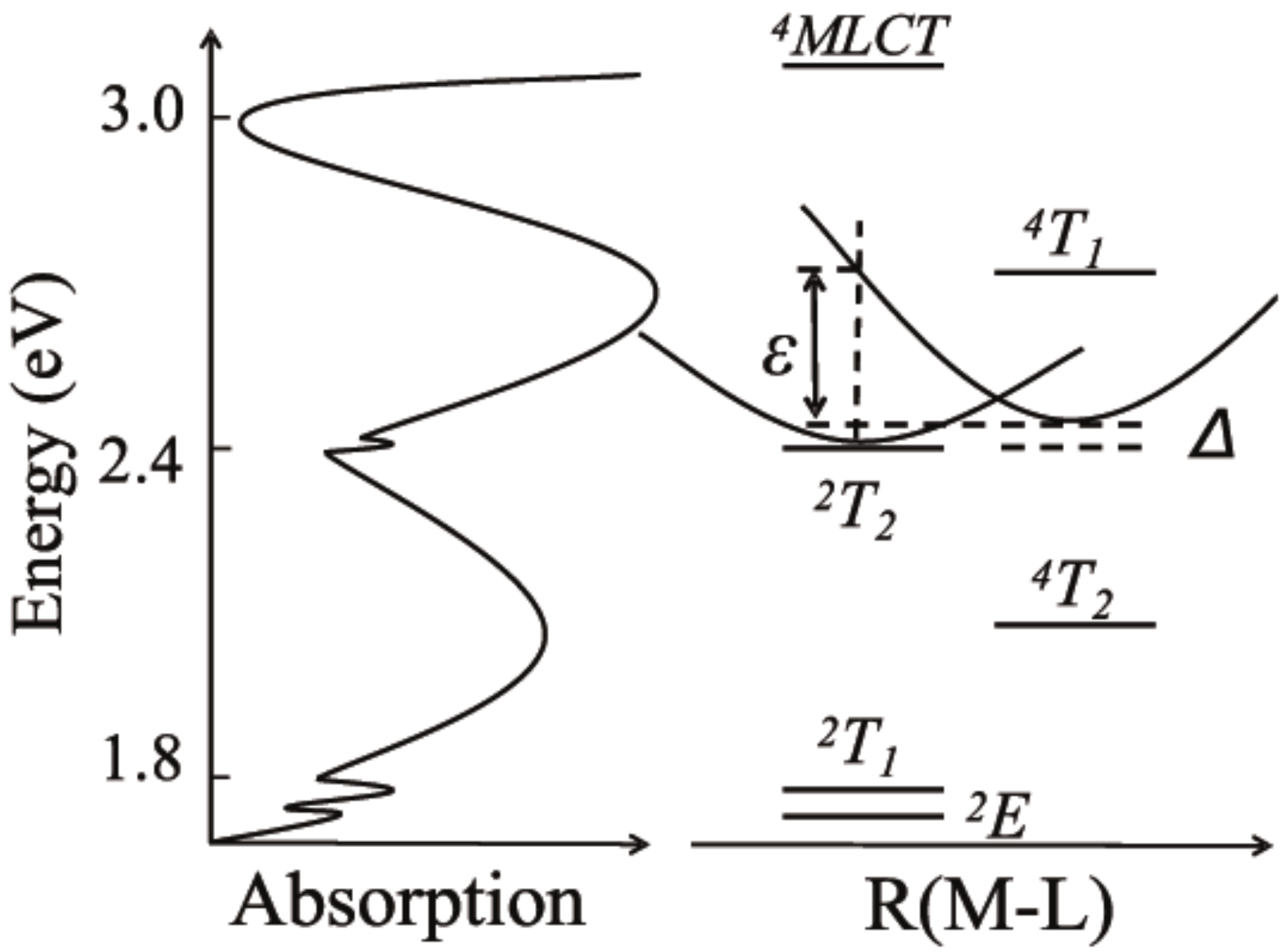} \caption{On the left: schematic absorbance of chromium oxide
based on the spectrum measurements found in Ref \protect\onlinecite{Mcclure1963Comparison}.
Two broad absorption bands located in the range of 1.8-3.0 eV correspond
to the spin-allowed transition from the $^{4}A_{2}$ ground state
to the $^{4}T_{2}$ and $^{4}T_{1}$ quartet states. The three sharp
lines are associated with the spin-forbidden transitions to the $^{2}T_{2}$,
$^{2}E$ and $^{2}T_{1}$ doublet states. On the right: energy-level
scheme of Cr$_{2}$O$_{3}$. $R$ represents the coordination along
the metal-ligand coordinate. $\Delta$ and $\varepsilon$ are the
energy gap between the lowest vibrational levels and the electron-phonon
self-energy difference between two oscillation states, respectively. The central energy value ${E_{i}}$ of the absorption spectrum is indicated by a line segment. In some cases, $E_{i}$ is different from the energy value of the lowest vibrational
level owing to different electron-phonon couplings, e.g. in the $^{4}T_{2}$ and $^{4}T_{1}$ states. Here, we have set the ground state energy to zero.}
\label{abs} 
\end{figure}

\section*{Demagnetization Model}

A typical static energy-level scheme of a Cr$^{3+}$ metal ion is
shown in Fig. \ref{abs}. The metal ion is in close proximity to oxygen octahedral surrounding
and the five-fold degenerate $3d$ orbitals are split into a lower
threefold-degenerate $t_{2g}$ and an upper twofold-degenerate $e_{g}$
orbitals by the crystal field with O$_{h}$ symmetry. Due to the Hund
coupling, the ground state, less than half filled, is a high-spin $\left(S=3/2\right)$$~^{4}A_{2}$
($t_{2g}^{3}$) configuration. Early in 1963, McClure reported the
polarized optical absorption spectrum of Cr$_{2}$O$_{3}$ with the
wavelength ranging from 300 to 800 nm in thin single-crystal plates. \cite{Mcclure1963Comparison}
Two broad absorption bands are observed in the range of 400\textendash 800
nm corresponding to the transitions of the $3d$ electron shell from the $^{4}A_{2}$ ground-state level to the excited-state levels,
$^{4}T_{1}$ ($t_{2g}^{2}$$e_{g}^{1}$) and $^{4}T_{2}$ ($t_{2g}^{2}e_{g}^{1}$),
respectively. \cite{Mo1997Modulation,Brik2004Crystal} Between
the $^{4}T_{2}$ and $^{4}T_{1}$ absorption bands, there is a sharp
line associated to the spin-forbidden transition to the $^{2}T_{2}$~$(t_{2g}^{3})$
doublet. Two other sharp lines link to the transitions to the
low-lying $^{2}E$~$(t_{2g}^{3})$ and $^{2}T_{1}$~$(t_{2g}^{3})$
doublets. \cite{Macfarlane1963Analysis,Krichevtsov1996Magnetoelectric,Ogasawara2016Multiplet,Torchia2004Phonon}

During the ultrafast photodirven demagnetization process from the high-spin to low-spin states in Cr oxides,
the first localized excited state triggered by laser irradiation does
not directly return to the ground state but follows a complex route
of intermediate states accompanying with changes in spin and lattice
parameters. The spin-orbit coupling (SOC) could flip the spin of $d$-orbit
electrons in the intermediate states. The redistribution of anisotropic
$d$-orbital occupations often leads to geometric deformation or structural
phase transition. Meanwhile, the locally excited state dissipates
energy to its surroundings by emission of phonons and/or
photons. Since the relaxation time of fluorescence is on a nanosecond (ns) time scale, then a phonon continuum dominates the energy dissipation
in the ultrafast demagnetization. To elucidate this dynamical process,
we introduce a model with electronic multiplet levels at energies
$E_{i}$, coupled to a phonon bath. Due to the strong electron-phonon
coupling and the substantial bath memory effects in a photodriven
system, a Born-Markov master equation fails to effectively describe the ultrafast
electron dynamics. Therefore, we first map the spin-boson-like model
to an alternative model, where the electronic levels are coupled to
a single harmonic mode damped by an Ohmic bath. \cite{Chang2014Model,Garg1985Effect}
The memory effects could be effectively taken into account by the
time evolution of the strength of the harmonic mode. Here, the correlations between electrons are taken into account by the renormalization of the electronic state energies. The local
system Hamiltonian is written as

\begin{eqnarray}
H_{S}=\sum_{i}E_{i}c_{i}^{\dagger}c_{i} & + & \hbar\omega a^{\dagger}a+\sum_{i}\lambda_{i}c_{i}^{\dagger}c_{i}(a^{\dagger}+a)\nonumber \\
 & + & \sum_{ij}V_{ij}(c_{i}^{\dagger}c_{j}+c_{j}^{\dagger}c_{i}),
\end{eqnarray}
where $c_{i}^{\dagger}c_{i}$ gives the occupation of the multiplet
$i$, $V_{ij}$ is the coupling constant that causes a transition
between energy level $j$ and $i$, $a^{+}$ is the creation operator
for the harmonic phonon with frequency $\omega$. We further define
the electron-phonon self-energy difference $\varepsilon_{ij}=(\lambda_{i}-\lambda_{j})^{2}/\hbar\omega$
and the energy gap $\Delta_{ij}=(E_{i}-\lambda_{i}^2/\hbar\omega)-(E_{j}-\lambda_{j}^2/\hbar\omega)$
between two states, where variations in the electron-phonon coupling
strength $\lambda_{i}$ change the equilibrium positions of different
states. (see Fig. \ref{abs}).

We describe the time evolution of the local open quantum system with
the dissipative Schr\"odinger equation \cite{Veenendaal10,Chang10}

\begin{eqnarray}
i\hbar\frac{d|\psi(t)>}{dt}=(H_{0}+iD)|\psi(t)>,
\end{eqnarray}
where $H_{0}$ is the the Fr\"ohlich transformation of the Hamiltonian
$H_{S}$, $D$ is a dissipative operator that describes the bath induced
state transfer \cite{Chang10}

\begin{eqnarray}
D=\frac{\hbar}{2}\sum_{k}\frac{d\mathrm{ln}P_{k}\text{\ensuremath{(t)}}}{dt}|\psi_{k}><\psi_{k}|.
\end{eqnarray}
Here, $P_{k}(t)=|c_{k}\left(t\right)|^{2}$ is the state probability and 
$|\psi(t)>=\sum_{k}c_{k}\left(t\right)|\psi_{k}>$. The time evolution
of the probability of multiplet $i$ with $n$ excited phonon modes
is given by \cite{Dekker1981Classical,Chang10}

\begin{eqnarray}
\frac{dP_{in}(t)}{dt}=-2n\Gamma P_{in}(t)+2(n+1)\Gamma P_{i,(n+1)}(t),
\end{eqnarray}
with $\Gamma=\pi\bar{\rho}\bar{V}^{2}/\hbar$, the environmental
phonon relaxation constant. \cite{Chang10,Chang12} According
to the Jablonski energy diagram, $(2\Gamma)^{-1}$ ranges from 0.01 picosecond (ps) to 10 ps. In this paper, we set $(2\Gamma)^{-1}=0.1$ ps.

\section*{demagnetization process}
Since the spin-flip is forbidden in photoexcitation, the first photoexcited
states starting from the $^{4}A_{2}$ ground state are $^{4}T_{2}$,
$^{4}T_{1}$, and the metal-ligand charge transfer $^{4}MLCT$ ($t_{2g}^{2}L^{1}$)
quartet states, depending on the energies of photoexciation. Here,
$L^{1}$ denotes that an electron transfers to ligands. After the light
illumination, the system decays to a relatively long-lived metastable
state, e.g., the $^{2}T_1$ and $^{2}E$ doublet states in Cr$_{2}$O$_{3}$.
Importantly, a long lifetime of the excited energy level promises
to be a candidate to develop a potential laser device. In order to
understand the dynamic process, we first need to determine the energies
of the states involved in the cascading process. According to the
absorption spectra, the central energies $E_{i}$ of the $^{2}E$, $^{2}T_{1}$ , $^{4}T_{2}$, $^{2}T_{2}$ , $^{4}T_{1}$,
and $^{4}MLCT$ states locate around 1.7, 1.76, 2.1, 2.45, 2.75, 3.3 eV,
respectively. \cite{Brik2005Crystal,Macfarlane1963Analysis,Mcclure1963Comparison}
Next, we need to establish an appropriate range of interaction parameters.
The strength of the interaction between the Cr ion and its surrounding
oxygen anions can be obtained from \textit{ab initio} calculations. \cite{Torchia2004Phonon}
The change in energy for different configurations is close to parabolic
for an adiabatic change in the Cr-ligand distance. From the change
in equilibrium distance or the optical absorption and luminescence
spectra, we can obtain the Huang-Rhys factor $g\approx4$ between
the doublets (e.g. $^{2}E$, $^{2}T_{1}$ and $^{2}T_{2}$) and the quartet
states (e.g. $^{4}T_{2}$ and $^{4}T_{1}$). \cite{Zundu1993} We assume the
Huang-Rhys factor $g\approx0$ between the ground state and $^{4}MLCT$ according to the sharp absorption line. Correspondingly, the difference
of electron-phonon self-energy $\varepsilon_{ij}$ is equal to $g\hbar\omega$
and the electron-phonon coupling constant $|\lambda_{i}-\lambda_{j}|=\sqrt{\varepsilon_{ij}\hbar\omega}$. The spin changes during the transfer from a quartet to a doublet state, and the coupling $V$ between the two different spin states is generally accepted to be due to the SOC. We take the strengh of SOC around 0.03 eV in Cr ions. \cite{Muto1998Magnetoelectric,Stamenova2016Role} Strongly coupled to the optically excited electrons, the optical phonon modes
could be observed by Raman spectroscopy. Owing to the symmetry of
Cr$_{2}$O$_{3}$, there are seven Raman modes, two with $A_{1g}$
symmetry and five with $E_{g}$ symmetry, and the longer wavelengths
corresponding to the $E_{g}$ modes. \cite{Sala2016Resonant} We take
the $E_{g}$ mode value $\hbar\omega=0.075$ eV, which dominates the
relaxation at the 1.8 eV pumping, and $A_{1g}$ mode $\hbar\omega=0.065$
eV, the main damping phonon at 3.0 eV photon excitation. \cite{Shim2004Raman}

\begin{figure}[t]
\includegraphics[width=1\columnwidth]{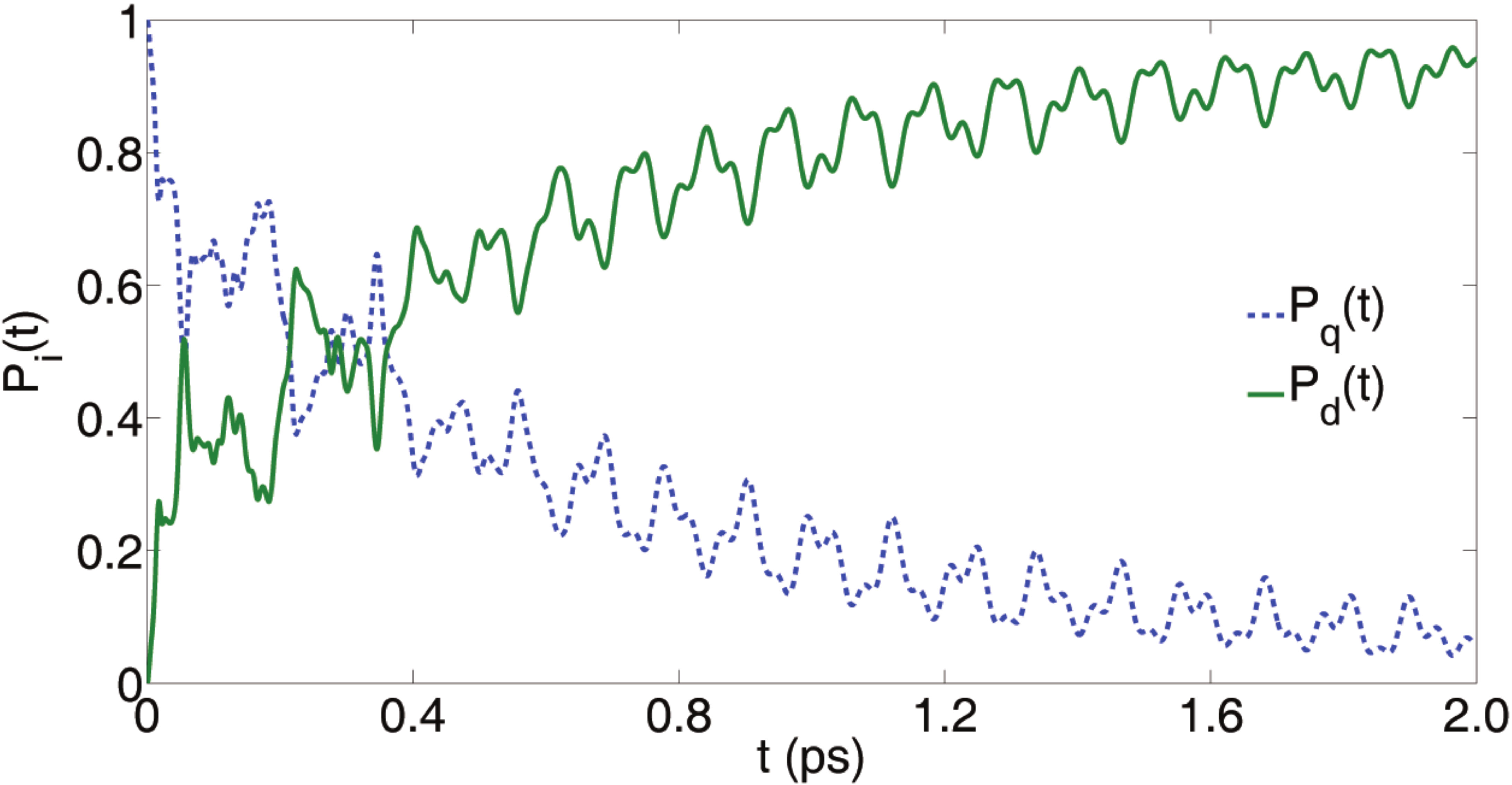} \caption{The probability of finding the quartet and doublet states as a function of time at 1.8 eV photon excitation.  The dashed line (blue) $P_{q}(t)$ gives the $^{4}T_{2}$ quartet state probability, the solid line (green) $P_{d}(t)$ shows the sum of the probability of the $^{2}T_1$ and $^{2}E$ doublet states. There are two oscillations in the state probabilities with the periods around 100 fs and 40 fs, corresponding to the 0.04 eV and 0.1 eV energy level gaps between two doublet states and the quartet state, respectively.}
\label{1.8eV} 
\end{figure}

The 1.8 eV photoexcitation results in the transition from the $^{4}A_{2}$
ground state to the $^{4}T_{2}$ excited state. An electron in the $t_{2g}$ orbital is locally excited to the $e_{g}$ orbital
by the illumination. Such a transition yields
an elongation of the Cr\textendash O band length of several tenths of an
{\AA}ngstrom since the change from a $t_{2g}$ to an $e_{g}$ charge
distribution leads to a stronger repulsion between the Cr and
the O ligands. \cite{Ogasawara2016Multiplet} The bond length change
leads to different electron-phonon couplings between the two states,
thereby forming a Franck-Condon continuum. \cite{Veenendaal10} Under
the action of SOC and electron phonon interaction,
the first excited state relaxes to the long lived states, namely,
the $^{2}T_{1}$ and $^{2}E$ doublet states. The energy gap between
$^{2}T_{1}$ and $^{2}E$ is small, e.g. around 0.06 eV, therefore
the $^{2}T_{1}$ and $^{2}E$ populations are often combined for kinetic
purposes. \cite{Forster2004Excited} In Fig. \ref{1.8eV}, we show
the time evolution of the three states involved in the ultrafast demagnetization
process by solving the dissipative Schr\"odinger equation. The starting state is $^{4}T_{2}$, excited from the ground state. The $E_{g}$ phonon mode with $\hbar\omega=0.075$ eV dominates the relaxation. \cite{Shim2004Raman} Therefore, the self-energy difference $\varepsilon$ between the metal-centered quartet and doublet is 0.3 eV, with the Huang-Rhys factor $g=4$. The quartet state and the doublet state are mediated by SOC, i.e. $V$=0.03 eV. The energy gap $\Delta$  between $^{4}T_{2}$ and $^{2}T_{1}$ is 0.04 eV, and 0.1 eV between $^{4}T_{2}$ and $^{2}E$.  The 0.04 eV and 0.1 eV energy gaps are indicated in the obvious oscillations with the periods around 100 fs and 40 fs, respectively in the time evolution of both the quartet and doublet states. We find that the probability of the $^{4}T_{2}$ state falls quickly and the sum of the probability of the  $^{2}T_{\text{1}}$ and $^{2}E$ states increases in the first
0.5 ps. Fitting the curves using kinetic rate equations, the rise
time constant of the sum of the two doublet states are around 400
fs, which agrees well with the experiments by the time-resolved
second harmonic generation. \cite{Sala2016Resonant}

\begin{figure}[t]
\includegraphics[width=1\columnwidth]{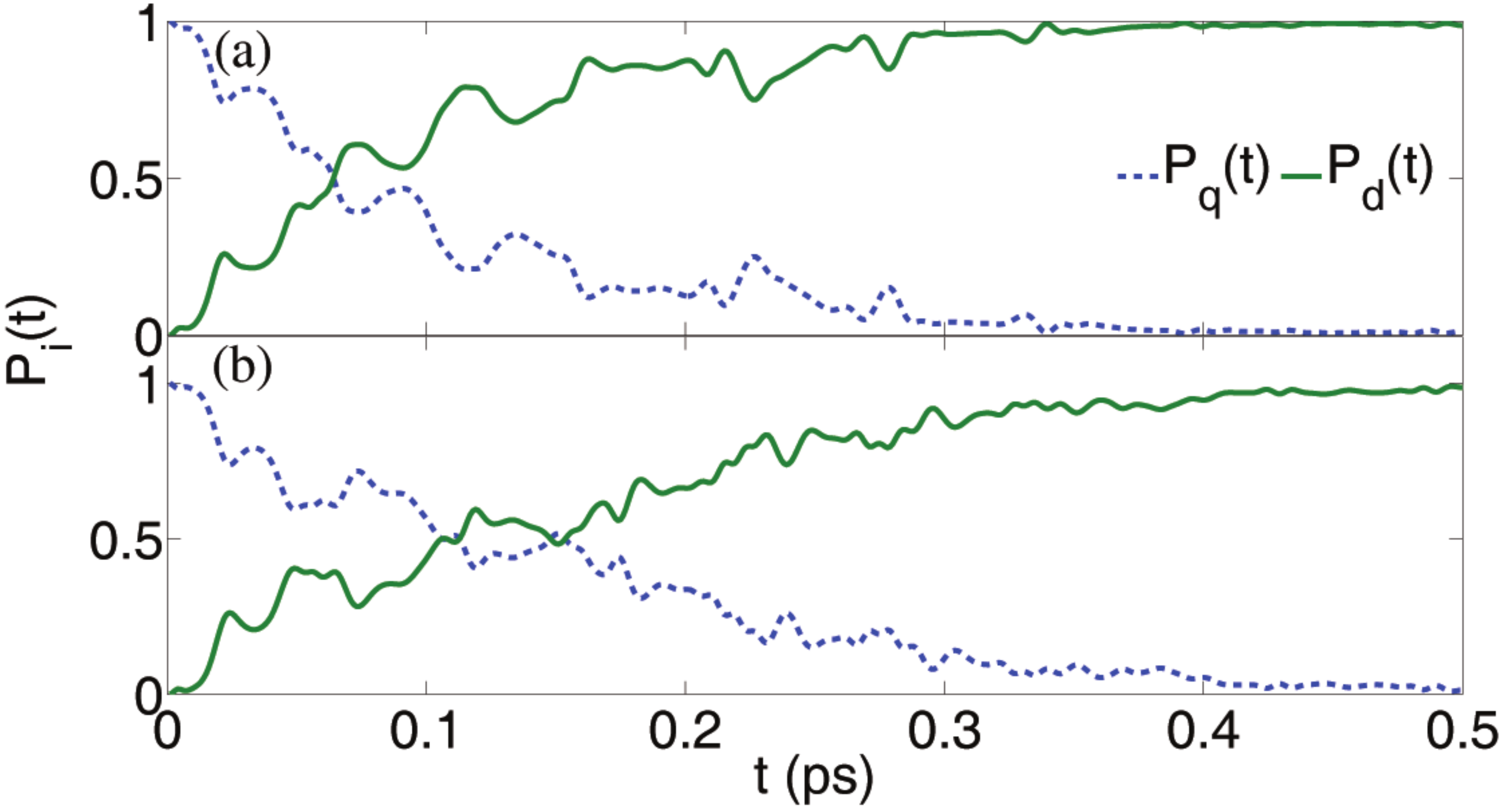} \caption{The time evolution of the state probabilities with different energy gaps between $^{4}T_{2}$ and $^{2}T_{1}$ at 1.8
eV photon pumping.  (a) The gap $\Delta$ between $^{4}T_{2}$
and $^{2}T_{1}$ is 0.24 eV, close to the corresponding $\varepsilon$=0.3 eV, the demagnetization time reduces to 100 fs; (b) $\Delta=0.34$
eV, the relaxation time is around 150 fs. The dashed line (blue) and
the solid line (green) give the probabilities of finding $^{4}T_{2}$
and the sum of $^{2}T_{1}$ and $^{2}E$, respectively.}

\label{ratio} 
\end{figure}

The decay time of a photoexcited state strongly depends on the ratio
of the energy gap to the electron-phonon self-energy difference, which
has been demonstrated in transition-metal complexes. \cite{Chang10}
When the ratio $\Delta/\varepsilon$
ranges from 0.5 to 1.5, the fastest decay occurs. In engineering,
the energy gap between the multiplets could be changed by distortion
stress, strain or chemical substitution of ligands, which provides
a feasible approach to adjust the demagnetization time. For example,
since the gap between $^{4}T_{2}$ and $^{2}T_{1}$ is very small, $\Delta=0.04$ eV, 
it may result in a longer decay time. We found that provided the
gap increases 0.2 eV, close to $\varepsilon=g\hbar\omega=0.3$ eV, the demagnetization time is strongly
reduced to around 100 fs, a quarter of the original period, as shown
in Fig. \ref{ratio}. Comparing with Fig. \ref{1.8eV}, the probability oscillations are strongly suppressed by the faster energy dissipation. On the other hand, if we only vary the photoexcitation energies from 1.8 eV to 2.1 eV and keep all the other parameters the same, the time evolutions of the three states change slightly from Fig. \ref{1.8eV} (not
shown).

\begin{figure}[t]
\includegraphics[width=1\columnwidth]{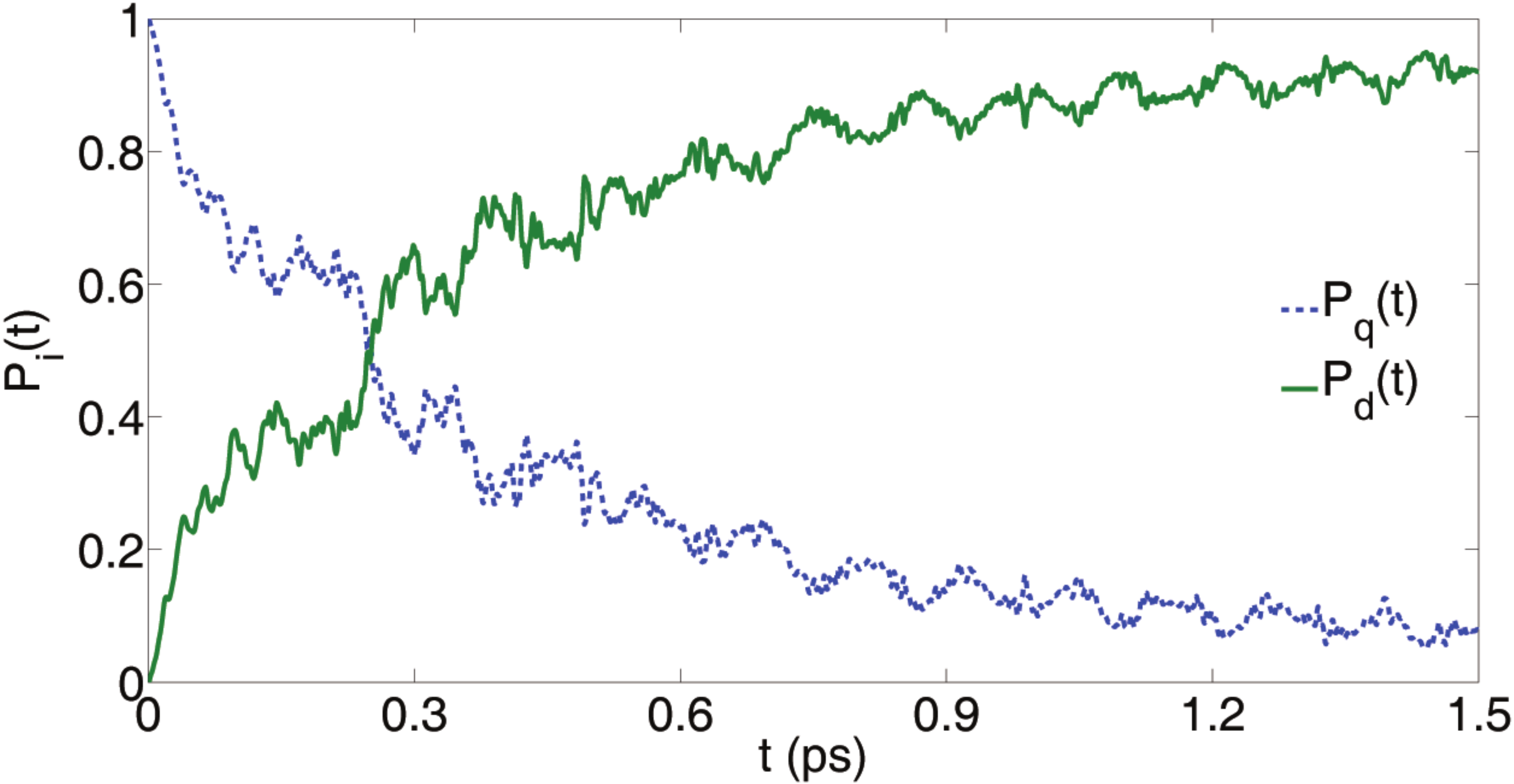} \caption{The electron occupation probability in demagnetization
process at 3.0 eV photon pumping. The dashed lines (blue) denotes
the time evolution of the sum of the $^{4}MLCT$ , $^{4}T_{1}$ and
$^{4}T_{2}$ quartet state probabilities, the solid line (green) refers
to the sum of the probability of the $^{2}T_{\text{2}}$ , $^{2}T_{\text{1}}$
and $^{2}E$ doublets. The first photoexcited state is $a\left|^{4}MLCT\right\rangle +\sqrt{1-a^{2}}\left|^{4}T_{1}\right\rangle $
with $a^{2}=0.25$. The fitting rise time value of the doublet probability is
around 300 fs. \label{mixed} }
\end{figure}

The 3.0 eV photon energy is supposed to excite the $^{4}A_{2}$ ground
state to the $^{4}T_{1}$ metal-centered state and/or the $^{4}MLCT$
metal-ligand charge transfer state. Due to the electron-phonon interaction,
SOC and orbital hybridization, the first excited state
finally reaches the $^{2}T_{1}$ and $^{2}E$ doublet states via the
transit $^{4}T_{1}$, $^{2}T_{2}$ and $^{4}T_{2}$ states. In pure
octahedral symmetry, there is no coupling between $^{4}MLCT$
and $^{4}T_{1,2}$, since the $e_{g}$ orbital states do not couple
to the ligand $\pi^{*}$ states. Nevertheless, since the nonequilibrilium
charge transfer often leads to a lower structural symmetry, a small
hybridization between the two quartet states should be presented,
depending on the amount of distortion. The weak hopping energy between
the ligands $\pi^{*}$ or $\pi$ and the metal ion's $e_{g}$ orbitals
is of the order of hundredths eV. \cite{Chang12} Our numerical calculations
are not sensitive to the change in the hybridization from 0.03 eV
to 0.09 eV. In Fig. \ref{mixed}, we set the hybridization parameter
0.05 eV between $\pi^{*}$ and $e_{g}$. The metal-centered quartet and doublet states are supposed to be mediated by SOC, $V$=0.03 eV. There is no direct coupling  between $^{4}MLCT$ and the doublet states since no interaction allows spin-flip and charge transfer synchronously. The main damping phonon is the $A_{1g}$ phonon mode with $\hbar\omega=0.065$ eV. \cite{Shim2004Raman} Consequently, $\varepsilon$ between the metal-centered quartet and doublet states is 0.26 eV with the Huang-Rhys factor $g=4$. First, we assume that the
electrons in the ground state is excited to the $^{4}MLCT$ state.  From the time evolution of the states, we find that the rise time constant of the doublets is around 360 fs by fitting
the curves using kinetic rate equations. Next, it has been pointed out
that the first excited state at high energy excitation could be the
mix of $^{4}T_{1}$ and $^{4}MLCT$. \cite{Zhang1997Characterizations}
Taking a mixed first excited state $a\left|^{4}MLCT\right\rangle +\sqrt{1-a^{2}}\left|^{4}T_{1}\right\rangle $,
the fitting rise time values of the doublets are 450 fs with the mix
constant $a^{2}=0.75$, 380 fs for $a^{2}=0.5$, and 300 fs for $a^{2}=0.25$,
respectively. In Fig. \ref{mixed}, the time-dependent occupations
of the states involved in the demagnetization process are shown, with
the mix constant $a^{2}=0.25$. Interestingly, the $^{2}T_{2}$ state
is often ignored in some literatures because its absorption is too
narrow to be resolved in optical spectra. However, if the $^{2}T_{2}$
state is omitted in our model, we find that the probability of $^{4}T_{1}$
plateaus at value 0.1 from 3 ps after the pumping. An extension of the time
evolution up to 0.1 ns confirms that the neglect of $^{2}T_{2}$
results in an incomplete decay of $^{4}T_{1}$, which is inconsistent with the experiments. Therefore, the time scale of ultrafast demagnetization depends not only on the pumping
energy, but also on the electric configuration of energy levels, the
spin-orbit and electron-phonon couplings.

\section*{Conclusion }

To conclude, we have presented a quantum-mechanical demagnetization
model for the locally photoinduced electron state in Cr$_{2}$O$_{3}$.
Using the dissipative Schr\"odinger equation, the environmental enegy dissipations are considered. We numerically simulated the time evolution
of the excited states following 1.8 eV and 3.0 eV photon excitation. The decay times are consistent with experiments on the order of hundreds of femtosecond from the high-spin quartet states
to the low-spin doublet states. Both the spin-orbit coupling and electron-phonon
coupling take important roles in the ultrafast demagnetization processes.
We have shown that the ratio of the energy gap to the electron-phonon self-energy
has strong impact on the demagnetization times. The hybridization of
the atomic orbitals in the first photoexcited state also affects the
decay times. We further expect that the demagnetization times could
be selectively controlled by the engineering stress-strain or chemical
substitution of ligands in insulating transition-metal oxides.

\section*{ACKNOWLEDGMENTS}

We are thankful to Jize Zhao,Hantao Lu and Ning Li  for fruitful discussions. F. G. and J. C. are
supported by the Fundamental Research Funds for the Central Universities,
Grant No. GK201402011. W. J. is supported by NSFC 11504223.

\end{document}